\pdfoutput=1
\documentclass[11pt,twoside,a4paper,cmspaper,final,collab]{cms-tdr}
\def\svnVersion{d55677e}\def\svnDate{2024/06/06}\def\cmsCernNoTag{CERN-EP-2024-058}\def\cmsCernDate{\today}\def\cmsMessage{Published in Physical Review D as \href{http://dx.doi.org/10.1103/PhysRevD.109.L111101}{\doi{10.1103/PhysRevD.109.L111101}.}}
\begin{document}\cmsNoteHeader{BPH-22-006}

\cmsNoteHeader{BPH-22-006}

\newcommand{\mumumumu}{\ensuremath{\PGmp\PGmm\PGmp\PGmm}\xspace}
\newcommand{\mumu}{\ensuremath{\PGmp\PGmm}\xspace}
\newcommand{\Jpsimumumumu}{\ensuremath{\JPsi\to\PGmp\PGmm\PGmp\PGmm}\xspace}
\newcommand{\Jpsimumu}{\ensuremath{\JPsi\to\PGmp\PGmm}\xspace}
\newcommand{\Mmumumumu}{\ensuremath{m_{\PGmp\PGmm\PGmp\PGmm}}\xspace}
\newcommand{\Mmumu}{\ensuremath{m_{\mumu}}\xspace}
\newcommand{\br}{\ensuremath{\mathcal{B}(\JPsi\to\mumumumu)}}
\newcommand{\result}{\ensuremath{\br=[10.1 ^{+3.3}_{-2.7}\stat\pm 0.4\syst] \times 10^{-7}}}
\newcommand{\brR}{\ensuremath{\mathcal{R}_{\Jpsimumumumu}}}
\newcommand{\resultR}{\ensuremath{\brR=[16.9 ^{+5.5}_{-4.6}\stat\pm 0.6\syst] \times 10^{-6}}}

\ifthenelse{\boolean{cms@external}}{\providecommand{\cmsLeft}{top\xspace}}{\providecommand{\cmsLeft}{left\xspace}}
\ifthenelse{\boolean{cms@external}}{\providecommand{\cmsRight}{bottom\xspace}}{\providecommand{\cmsRight}{right\xspace}}
\newlength{\cmsFigSize}
\ifthenelse{\boolean{cms@external}}{\setlength{\cmsFigSize}{0.30\textwidth}}{\setlength{\cmsFigSize}{0.45\textwidth}}

\title{\texorpdfstring{Observation of the \Jpsimumumumu decay in proton-proton collisions at $\sqrt{s} = 13\TeV$}{Observation of the J/psi to mu+mu-mu+mu- decay in proton-proton collisions at sqrt(s) = 13 TeV}}

\date{\today}

\abstract{The \Jpsimumumumu decay has been observed with a statistical significance in excess of five standard deviations. The analysis is based on an event sample of proton-proton collisions at a center-of-mass energy of 13\TeV, collected by the CMS experiment in 2018 and corresponding to an integrated luminosity of 33.6\fbinv. Normalizing to the \Jpsimumu decay mode leads to a branching fraction of $[10.1^{+3.3}_{-2.7}\stat \pm0.4\syst]\times 10^{-7}$, a value that is consistent with the standard model prediction.}

\hypersetup{%
pdfauthor={CMS Collaboration},%
pdftitle={Observation of the J/psi to mu+mu-mu+mu- decay in in proton-proton collisions at sqrt(s) = 13 TeV},%
pdfsubject={CMS},%
pdfkeywords={flavor, rare decays, four muons}}

\maketitle

Decays of particles to leptons, especially muons, 
provide some of the cleanest signatures at hadron collider experiments.
The large data sample collected by the CMS experiment~\cite{CMS:2008xjf,CMS:2309.05466} 
at the CERN LHC offers an excellent opportunity to explore rare decays to multilepton final states. 
The first such observation was the decay of the \PZ boson into four leptons~\cite{CMS:2012bw}. 
Subsequently, the decays $\PZ \to \JPsi\,\Pell^{+}\Pell^{-}$ 
(with the \JPsi decaying to two muons) and $\eta\to\mumumumu$ 
were also observed~\cite{CMS:2018vcg,CMS:2023thf}.

\begin{figure}[ht!]
\centering
\includegraphics[width=\cmsFigSize]{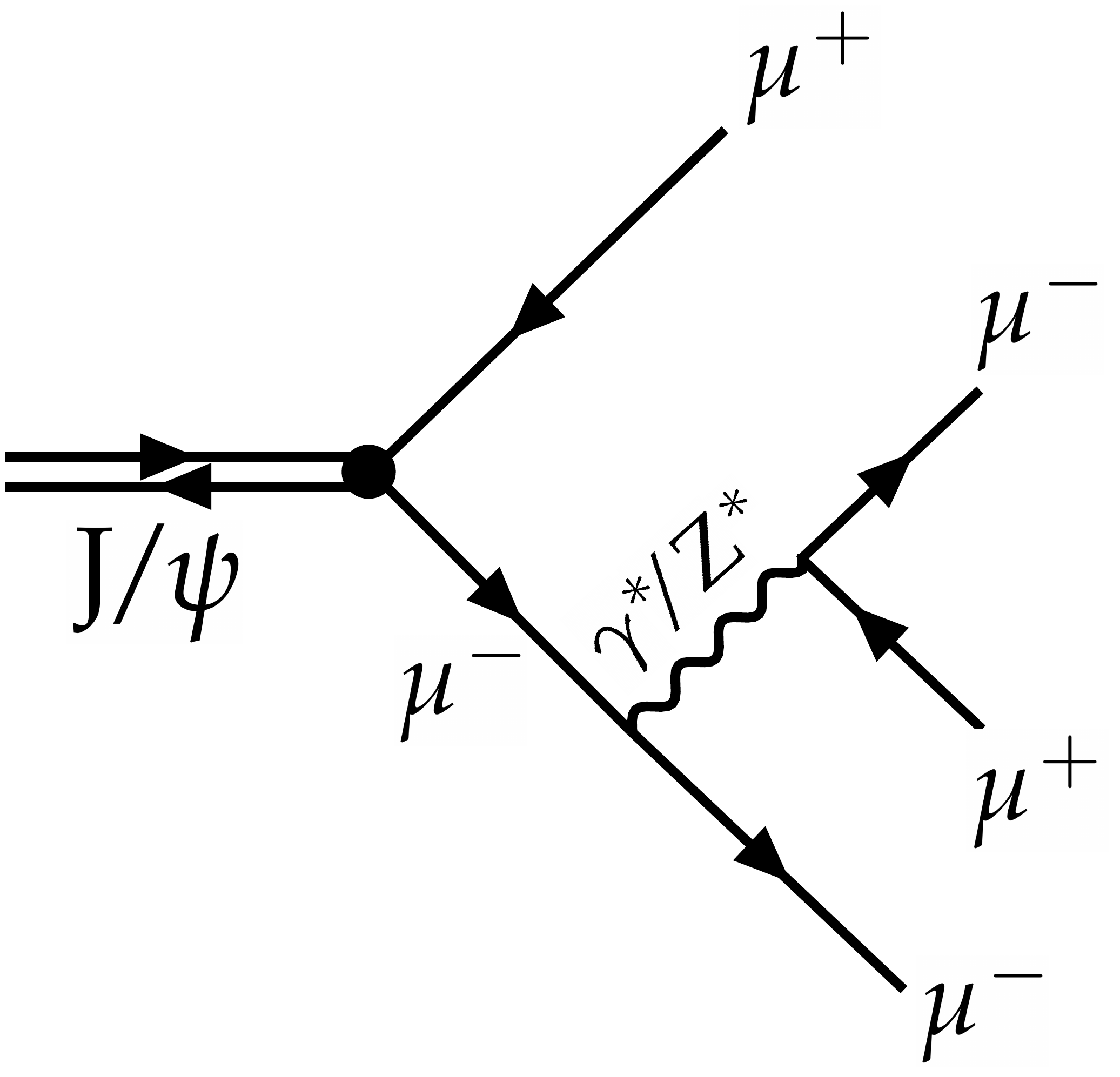}
\caption{Leading-order Feynman diagram representing the \Jpsimumumumu decay channel.\label{fig:diag}}
\end{figure}

The BES\,III Collaboration recently reported the observation of the \JPsi decays 
$\JPsi \to \Pep\Pem\Pep\Pem$ and $\JPsi \to \Pep\Pem\PGmp\PGmm$,
while for the \Jpsimumumumu branching fraction an upper limit was established, 
$1.6\times 10^{-6}$ at 90\% confidence level~\cite{BESIII:2021ocn}. 
In the standard model (SM), 
these processes occur via $\Pell \to \Pell \, \PGg^{\ast}/\PZ^{\ast}$ transitions, 
where the virtual photon or \PZ boson decays into a pair of leptons, 
as depicted in Fig.~\ref{fig:diag}. 
These transitions provide opportunities to probe various beyond-SM scenarios,
where new particles replace the $\PGg^{\ast}$ or the $\PZ^{\ast}$ boson~\cite{Drees:2018hhs,Harigaya:2013twa,Bell:2014tta}. 
Furthermore, such rare multilepton decays serve as a novel testing ground 
for quantum electrodynamics predictions~\cite{Chen:2020bju,Gutschow:2020cug}.

This Letter presents the first observation of the \Jpsimumumumu decay 
and the measurement of its branching fraction, relative to the \Jpsimumu decay mode.
The analysis is based on a sample of proton-proton collisions 
at a center-of-mass energy of 13\TeV,
collected by the CMS experiment
and corresponding to an integrated luminosity of 33.6\fbinv~\cite{CMS-PAS-LUM-18-002}.

The central feature of the CMS apparatus is a superconducting solenoid of 6\unit{m} internal diameter, 
providing a magnetic field of 3.8\unit{T}. 
Within the solenoid volume are a silicon pixel and strip tracker, 
a lead tungstate crystal electromagnetic calorimeter, 
and a brass and scintillator hadron calorimeter, 
each composed of a barrel and two endcap sections. 
Forward calorimeters extend the pseudorapidity ($\eta$) coverage 
provided by the barrel and endcap detectors. 
Muons are measured in the $\abs{\eta} < 2.4$ range, 
with detection planes made using three technologies: 
drift tubes, cathode strip chambers, and resistive plate chambers. 
A more detailed description of the CMS detector can be found in Ref.~\cite{CMS:2008xjf}.

Events of interest are selected using a two-tiered trigger system. 
The first level, composed of custom hardware processors, 
uses information from the calorimeters and muon detectors 
to select events at a rate of around 100\unit{kHz} within a fixed latency of 4\mus~\cite{CMS:2020cmk}. 
The second level
consists of a farm of processors running a version of the full event reconstruction software 
optimized for fast processing
and reduces the event rate before data storage~\cite{CMS:2016ngn}.

The single-muon trigger efficiency exceeds 90\% over the full $\eta$ range, 
and the efficiency to reconstruct and identify muons is larger than 96\%. 
By matching muons to tracks measured in the silicon tracker, 
the transverse momentum, \pt, 
is measured with a relative resolution of 1\% in the barrel and 3\% in the endcaps, 
for muons with \pt up to 100\GeV~\cite{CMS:2018rym}. 

This study exploits the ``\PB Parking" data sample collected by the CMS experiment in 2018~\cite{CMS:2024zhe}. 
A specialized trigger and data storage strategy was implemented to assemble a data set 
enriched in \PQb hadron decays~\cite{CMS:2024syx}. 
The trigger selects events with at least one muon 
with $\pt>9\GeV$ and transverse impact parameter significance
(distance of closest approach of the track to the beam line~\cite{CMS:2014pgm} 
divided by its uncertainty) larger than 6.

Two simulated event samples, used to evaluate the efficiency to reconstruct and select the 
\Jpsimumumumu and \Jpsimumu decays, are generated with the \PYTHIA~8.230 Monte Carlo event 
generator~\cite{Sjostrand:2014zea}, which includes modeling of the parton shower, fragmentation, 
and hadronization processes. The \PYTHIA output is interfaced with \EVTGEN~1.3.0~\cite{Lange:2001uf}, 
which simulates various \PQb hadron decays. The samples are generated such that each event 
contains at least one \PQb hadron decaying to a \JPsi meson plus other decay products, with the 
\JPsi meson decaying to either two or four muons using a phase space model. The 
underlying event is also modeled with \PYTHIA8, adopting the CP5 tune~\cite{CMS:2019csb}. 
The NNPDF~3.1~\cite{NNPDF:2017mvq} parton distribution functions are used. To account for 
additional proton-proton collisions in the same or adjacent bunch crossing (pileup), simulated minimum 
bias events are superimposed onto the hard scattering process, matching the multiplicity of 
reconstructed vertices observed in collision data.
Final-state photon radiation is simulated using \PHOTOS~3.61~\cite{Barberio:1993qi}. 
The response of the CMS detector is emulated using \GEANTfour~\cite{Agostinelli:2002hh}. 
The simulated event samples are reconstructed using the same software packages as employed for collision data.

The events selected by the trigger need to comply with additional selection criteria, 
including the presence of four reconstructed muons with a net charge of zero 
and an invariant mass in the $2.6 < \Mmumumumu < 3.4\GeV$ range. 
All the four selected muons are required to have $\abs{\eta} < 1.5$ and $\pt > 3.5\GeV$. 
One of them must match the muon that triggered the event, 
which is ensured by requiring that it has $\pt > 9\GeV$ 
and a small angular distance with respect to the trigger muon,
$\sqrt{\smash[b]{\Delta\eta^2 + \Delta\phi^2}} < 0.1$,
where $\Delta\eta$ and  $\Delta\phi$ are 
the differences in pseudorapidity and azimuthal angle, respectively.

The four muons forming a \JPsi candidate are fitted to a common vertex, 
determined by a Kalman filter vertex fit~\cite{Fruhwirth:1987fm}.
Only candidates with a vertex-fit $\chi^2$ probability exceeding 1\% are kept.
The \pt of the \JPsi meson candidate must exceed 25\GeV 
and its absolute rapidity is restricted to be less than 2.2. 
No oppositely-charged muon pair combination can have an invariant mass in the 0.75--0.80 or 0.98--1.05\GeV ranges, 
which are populated by dimuon decays of the $\PGr$, $\PGo$, and $\PGf$ mesons.

The branching fraction of the \Jpsimumumumu decay mode 
is measured relative to that of the \Jpsimumu decay. 
The selection criteria used in the \Jpsimumu events are the same as those
applied to the two highest \pt muons of the \Jpsimumumumu decay.

\begin{figure}[th!]
\centering
\ifthenelse{\boolean{cms@external}}
{\includegraphics[width=\linewidth]{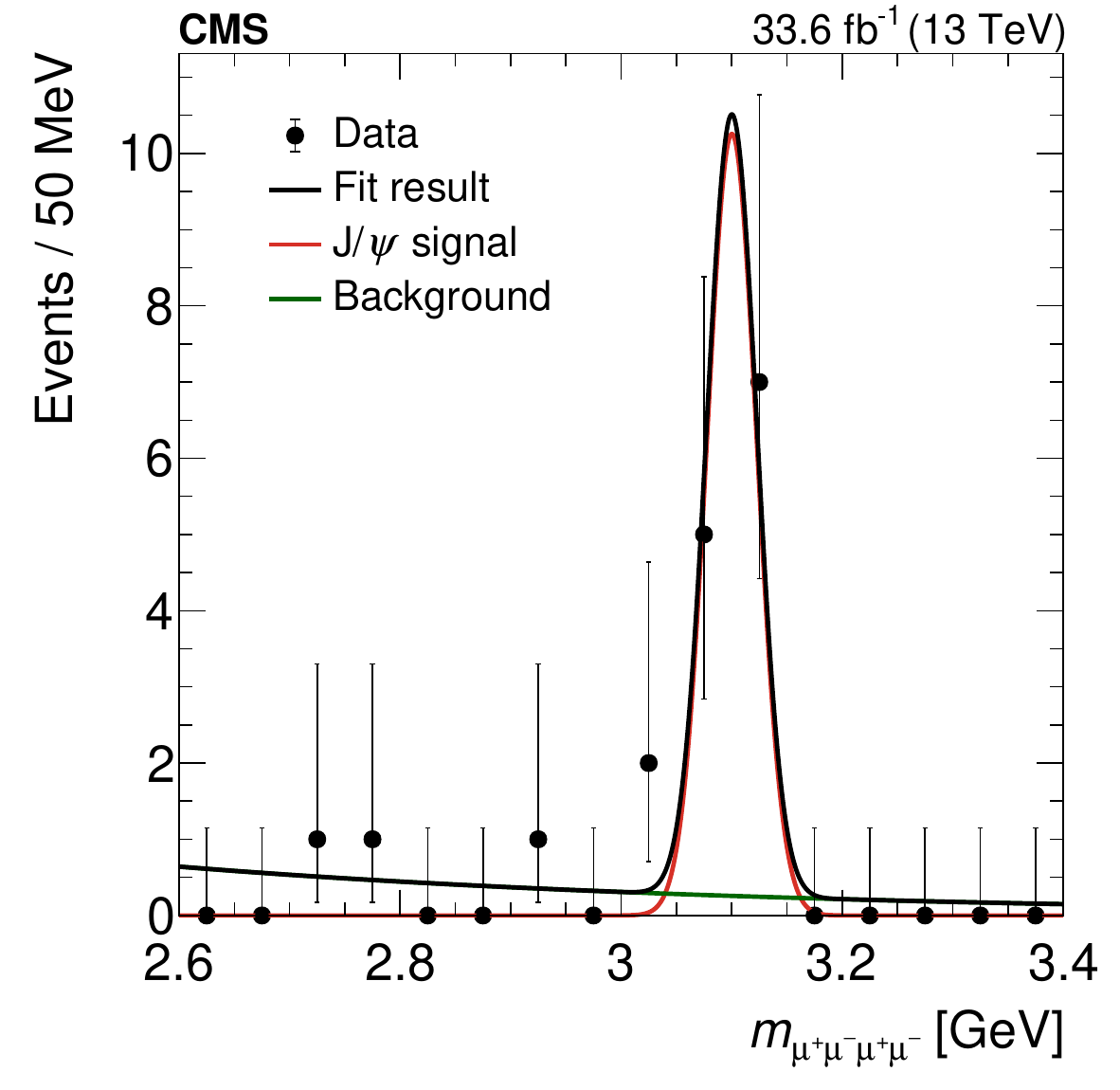}\\
\includegraphics[width=\linewidth]{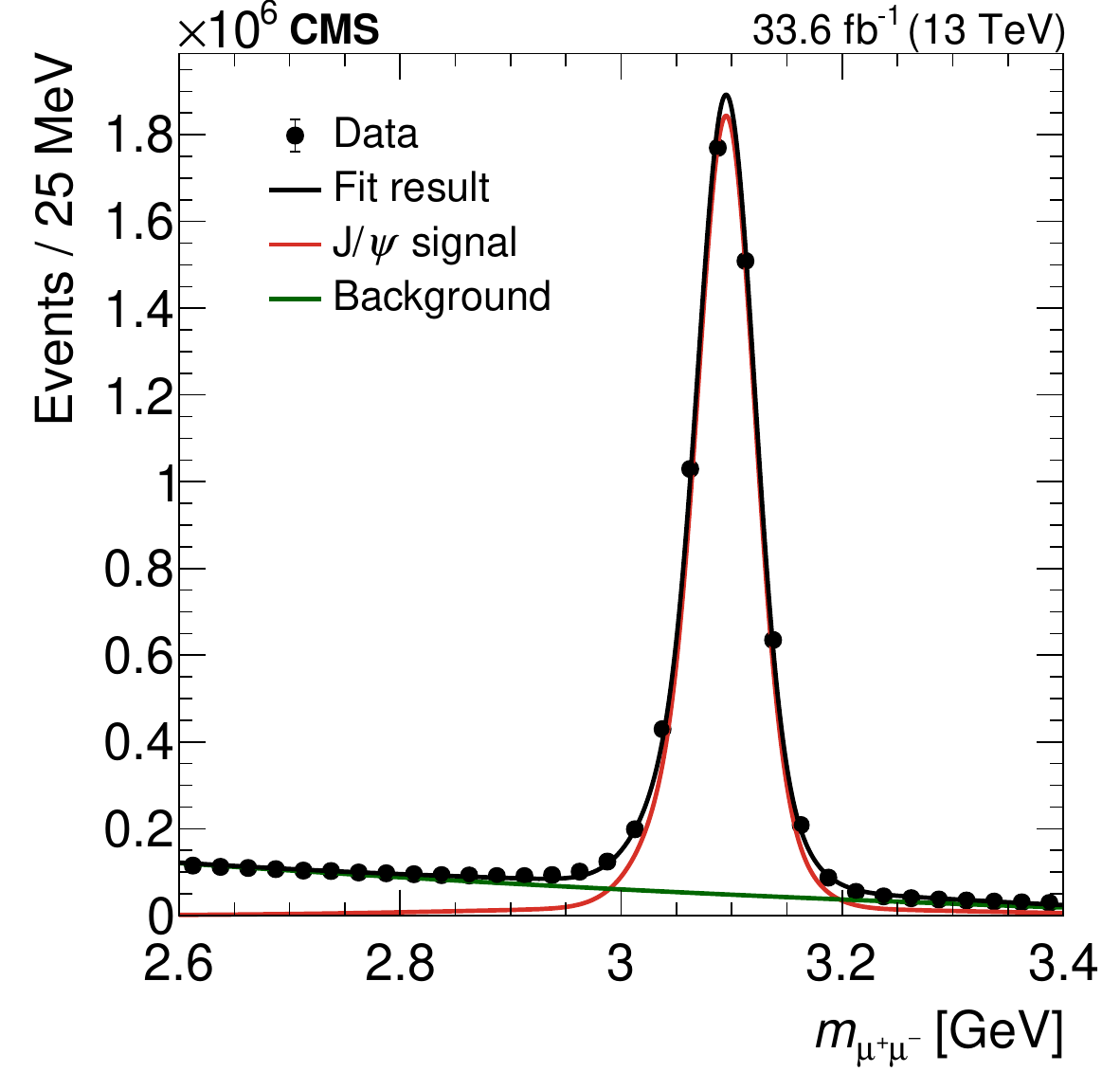}}
{\includegraphics[width=0.49\linewidth]{Figure_002-a.pdf}
\includegraphics[width=0.49\linewidth]{Figure_002-b.pdf}}
\caption{Measured four-muon (\cmsLeft) and dimuon (\cmsRight) mass distributions.
The vertical bars represent the statistical uncertainties. 
The solid black line represents the result of the unbinned fit described in the text, 
while the red and green lines represent the signal and background terms, respectively.}
\label{fig:massbr}
\end{figure}

Unbinned maximum likelihood fits are performed on the four-muon and dimuon invariant mass distributions 
to determine the two signal yields. 
The \mumumumu invariant mass distribution is fitted by the superposition of two functions,
one representing the \Jpsimumumumu signal and the other the underlying continuum background.
The signal is modeled by a Crystal Ball function~\cite{CrystalBallRef},
with the mean fixed to the world-average \JPsi meson mass~\cite{ParticleDataGroup:2020ssz}
and the width and tail parameters fixed to values determined from studies of simulated events.
The background is modeled by a linear function.
The \mumu invariant mass distribution is fitted with essentially the same fit model,
except that the signal is represented by the sum of a Crystal Ball function 
and a Gaussian function, without constraining any of the parameters.
The fit models do not include peaking background terms from hadronic \JPsi decay modes
because the probability that a pion reaches the muon stations 
and is misidentified as a muon is below the per mil level~\cite{CMS:2018rym}, 
so that such contributions are negligible.
The measured \Mmumumumu and \Mmumu distributions are shown in Fig.~\ref{fig:massbr},
together with the results of the fits.

The yields returned by the fits are $N(\Jpsimumumumu) = 11.6^{+3.8}_{-3.1}$ 
and $N(\JPsi\to\mumu) = (5770 \pm 3)\times 10^{3}$. 
The significance of the \Jpsimumumumu signal is above 7 standard deviations, evaluated from 
the likelihood ratio of the default signal-plus-background fit and the background-only 
fit, imposing $N(\Jpsimumumumu) = 0$, using the standard asymptotic formula~\cite{Cowan:2010js}.

The \Jpsimumumumu branching fraction 
relative to that of the \Jpsimumu is computed as
\ifthenelse{\boolean{cms@external}}
{
\begin{multline}
 \frac{\br}{\mathcal{B}(\JPsi\to\mumu)} \\
= \frac{N(\JPsi\to\mumumumu)}{N(\JPsi\to\mumu)} \, \Big/ \,
\frac{\epsilon_{\Jpsimumumumu}}{\epsilon_{\Jpsimumu}} ,
\label{eq:formula}
\end{multline}
}
{
\begin{equation}
\frac{\br}{\mathcal{B}(\JPsi\to\mumu)} =  
\frac{N(\JPsi\to\mumumumu)}{N(\JPsi\to\mumu)} \, \Big/ \,
\frac{\epsilon_{\Jpsimumumumu}}{\epsilon_{\Jpsimumu}} ,
\label{eq:formula}
\end{equation}
}
where $\mathcal{B}(\JPsi\to\mumu)$ is 
the branching fraction of the reference channel. 
The reconstruction efficiencies of each process  
are calculated as the fractions of generated events that are reconstructed,
their ratio being $\epsilon_{\Jpsimumumumu} / \epsilon_{\Jpsimumu} = (11.92 \pm 0.02)\%$, 
where the uncertainty reflects the size of the simulated samples. 

Because the signal- and reference-channel events are recorded using the same trigger 
and share similar event topologies, 
many systematic uncertainties have been seen to cancel in Eq.~(\ref{eq:formula}). 
In the following, we only describe those that do not cancel.

We evaluate the sensitivity of $N(\Jpsimumumumu)$ to the fit model by 
replacing the Crystal Ball function by a Gaussian function
and the linear function by an exponential function. 
The corresponding systematic uncertainties are 
smaller than 0.1\% when we replace the signal fit model 
and 0.4\% when we replace the background model. 
For the alternative modeling of the reference channel 
we use a sum of two Crystal Ball functions for the signal and 
an exponential function for the background. The fitted value of 
$N(\Jpsimumu)$ changes by 0.5\% for each of the two variations. 
An additional 0.1\% uncertainty reflects the size of the Monte Carlo samples 
used for calculating the reconstruction and selection efficiencies of the signal and reference channels.
Differences between the reconstructed and simulated samples are accounted for through 
two scale factors, one reflecting the trigger efficiency
and the other the reconstruction and event selection efficiencies.
Varying these scale factors by their standard deviation uncertainties,
we obtain the corresponding uncertainties on the branching fraction measurement:
1.9\% and 1.4\%, respectively. 
Recomputing the detection efficiencies with conservative variations 
of the simulated \JPsi kinematical distributions leads to a systematic uncertainty of 2.8\%. 
The total relative systematic uncertainty is 3.7\%,
computed as the quadratic sum of the individual terms,
so that the final result is
\ifthenelse{\boolean{cms@external}}
{
\begin{multline*}
\frac{\br}{\mathcal{B}(\JPsi\to\mumu)} \\
= [16.9 ^{+5.5}_{-4.6}\stat\pm 0.6\syst] \times 10^{-6}.
\end{multline*}
}
{
\begin{equation*}    
\frac{\br}{\mathcal{B}(\JPsi\to\mumu)} = 
[16.9 ^{+5.5}_{-4.6}\stat\pm 0.6\syst] \times 10^{-6}. 
\end{equation*}
}

Using the world's average 
$\mathcal{B}(\JPsi\to\mumu) = (5.961\pm 0.033)\%$~\cite{ParticleDataGroup:2020ssz},
and propagating its uncertainty, we obtain

\ifthenelse{\boolean{cms@external}}
{
\begin{multline*}
\br  \\
= [10.1 ^{+3.3}_{-2.7}\stat\pm 0.4\syst] \times 10^{-7},
\end{multline*}
}
{
\begin{equation*}
\br = [10.1 ^{+3.3}_{-2.7}\stat\pm 0.4\syst] \times 10^{-7},
\end{equation*}
}
a value consistent with the SM prediction, $(9.74\pm 0.05)\times 10^{-7}$~\cite{Chen:2020bju}.

The results reported in this paper are tabulated in the HEPData record for this analysis~\cite{hepdata}.

In summary, using a data sample of proton-proton collisions at $\sqrt{s} = 13\TeV$,
collected in 2018 by the CMS Collaboration and corresponding to an integrated luminosity of 33.6\fbinv,
the \Jpsimumumumu decay was observed for the first time, 
with a statistical significance exceeding five standard deviations. 
Taking the \Jpsimumu decay mode as normalization,
its branching fraction was measured to be \result, 
a value consistent with the standard model prediction.

\begin{acknowledgments}
We congratulate our colleagues in the CERN accelerator departments for the excellent performance of the LHC and thank the technical and administrative staffs at CERN and at other CMS institutes for their contributions to the success of the CMS effort. In addition, we gratefully acknowledge the computing centers and personnel of the Worldwide LHC Computing Grid and other centers for delivering so effectively the computing infrastructure essential to our analyses. Finally, we acknowledge the enduring support for the construction and operation of the LHC, the CMS detector, and the supporting computing infrastructure provided by the following funding agencies: SC (Armenia), BMBWF and FWF (Austria); FNRS and FWO (Belgium); CNPq, CAPES, FAPERJ, FAPERGS, and FAPESP (Brazil); MES and BNSF (Bulgaria); CERN; CAS, MoST, and NSFC (China); MINCIENCIAS (Colombia); MSES and CSF (Croatia); RIF (Cyprus); SENESCYT (Ecuador); ERC PRG, RVTT3 and MoER TK202 (Estonia); Academy of Finland, MEC, and HIP (Finland); CEA and CNRS/IN2P3 (France); SRNSF (Georgia); BMBF, DFG, and HGF (Germany); GSRI (Greece); NKFIH (Hungary); DAE and DST (India); IPM (Iran); SFI (Ireland); INFN (Italy); MSIP and NRF (Republic of Korea); MES (Latvia); LMTLT (Lithuania); MOE and UM (Malaysia); BUAP, CINVESTAV, CONACYT, LNS, SEP, and UASLP-FAI (Mexico); MOS (Montenegro); MBIE (New Zealand); PAEC (Pakistan); MES and NSC (Poland); FCT (Portugal); MESTD (Serbia); MCIN/AEI and PCTI (Spain); MOSTR (Sri Lanka); Swiss Funding Agencies (Switzerland); MST (Taipei); MHESI and NSTDA (Thailand); TUBITAK and TENMAK (Turkey); NASU (Ukraine); STFC (United Kingdom); DOE and NSF (USA).
\end{acknowledgments}

\bibliography{auto_generated}
\cleardoublepage \appendix\section{The CMS Collaboration \label{app:collab}}\begin{sloppypar}\hyphenpenalty=5000\widowpenalty=500\clubpenalty=5000
\cmsinstitute{Yerevan Physics Institute, Yerevan, Armenia}
{\tolerance=6000
A.~Hayrapetyan, A.~Tumasyan\cmsAuthorMark{1}\cmsorcid{0009-0000-0684-6742}
\par}
\cmsinstitute{Institut f\"{u}r Hochenergiephysik, Vienna, Austria}
{\tolerance=6000
W.~Adam\cmsorcid{0000-0001-9099-4341}, J.W.~Andrejkovic, T.~Bergauer\cmsorcid{0000-0002-5786-0293}, S.~Chatterjee\cmsorcid{0000-0003-2660-0349}, K.~Damanakis\cmsorcid{0000-0001-5389-2872}, M.~Dragicevic\cmsorcid{0000-0003-1967-6783}, P.S.~Hussain\cmsorcid{0000-0002-4825-5278}, M.~Jeitler\cmsAuthorMark{2}\cmsorcid{0000-0002-5141-9560}, N.~Krammer\cmsorcid{0000-0002-0548-0985}, A.~Li\cmsorcid{0000-0002-4547-116X}, D.~Liko\cmsorcid{0000-0002-3380-473X}, I.~Mikulec\cmsorcid{0000-0003-0385-2746}, J.~Schieck\cmsAuthorMark{2}\cmsorcid{0000-0002-1058-8093}, R.~Sch\"{o}fbeck\cmsorcid{0000-0002-2332-8784}, D.~Schwarz\cmsorcid{0000-0002-3821-7331}, M.~Sonawane\cmsorcid{0000-0003-0510-7010}, S.~Templ\cmsorcid{0000-0003-3137-5692}, W.~Waltenberger\cmsorcid{0000-0002-6215-7228}, C.-E.~Wulz\cmsAuthorMark{2}\cmsorcid{0000-0001-9226-5812}
% ...
\par}
\vskip\cmsinstskip
\dag:~Deceased\\
$^{1}$Also at Yerevan State University, Yerevan, Armenia\\
$^{2}$Also at TU Wien, Vienna, Austria\\
% ...
\end{sloppypar}
\end{document}